\documentstyle[11pt,newpasp,twoside]{article}
\def\edcomment#1{\iffalse\marginpar{\raggedright\sl#1\/}\else\relax\fi}
\reversemarginpar

\begin{document}
\title{Dark Matter Constraints from the Sagittarius Dwarf and Tail System}
 \author{Steven R. Majewski, David R. Law}
\affil{University of Virginia, Dept. of Astronomy, P.O. Box 3818, Charlottesville, VA 22903-0818, USA}
\author{Kathryn V. Johnston}
\affil{Wesleyan University, Dept. of Astronomy, Middletown, CT 06459-0123, USA}
\author{Michael F. Skrutskie}
\affil{University of Virginia, Dept. of Astronomy, P.O. Box 3818, Charlottesville, VA 22903-0818, USA}
\author{Martin D. Weinberg}
\affil{University of Massachusetts, Dept. of Physics \& Astronomy, 517 Lederle GRC, Amherst, MA 01003 USA}

\begin{abstract}
2MASS has provided a three-dimensional 
map of the $>360^{\circ}$, wrapped tidal tails of the 
Sagittarius (Sgr) dwarf spheroidal galaxy, as traced by M 
giant stars.  With the inclusion of radial velocity data 
for stars along these 
tails, strong constraints exist for dynamical models 
of the Milky Way-Sgr interaction.  N-body simulations of
Sgr disruption with model parameters spanning a range 
of initial conditions (e.g., Sgr mass and orbit, Galactic 
rotation curve, halo flattening)
are used to find parameterizations that match 
almost every extant observational constraint of the Sgr system.  
We discuss the implications of the Sgr data and models for 
the orbit, mass and $M/L$ of the Sgr bound core as well as the 
strength, flattening, and lumpiness of the Milky Way potential.
\end{abstract}

\section{Observational Constraints}

The relatively nearby Sagittarius (Sgr) dwarf galaxy offers the opportunity to explore in
exquisite detail the interaction of a satellite with its parent galaxy.  Moreover, the extensive
Sgr tidal tail system gives sensitive leverage on the properties of the Galactic potential, much
as polar ring galaxies have been exploited to determine the properties of extragalactic systems
(e.g., Sparke 2002).  

Since Sgr's discovery by Ibata, Gilmore \& Irwin (1994), this archetype of a dwarf galaxy merger
has remained difficult to study because its core (centered at $[l,b]=[6,-14]^{\circ}$),
lies obscured by foreground dust and stars of the Galactic disk and bulge.  However, the Two 
Micron All Sky Survey (2MASS) has opened a new window on Sgr because of the reduced extinction
by dust in the near infrared (NIR) and because Sgr contains a significant population 
of M giant stars, which are bright in the NIR.  To improve the contrast of
Sgr, these M giants can be separated from foreground M dwarfs using the combination of
2MASS $JHK_s$ photometry because of differential, surface gravity-sensitive opacity effects 
near 1.6$\mu$ (Bessell \& Brett 1988).  
Majewski et al. (2003a; MSWO hereafter) used this method to make
all-sky maps that reveal Sgr to be the primary source ($>80\%$) of M giants in the high halo
($|Z_{GC}|>20$ kpc), excluding those bound in the LMC and SMC (which actually show no M giant
tidal tails themselves).  The Sgr M giants lie along 
a great circle tipped only 13.5$^{\circ}$ from a truly polar ring.

A three dimensional analysis is made possible by converting $K_s$ magnitudes into
photometric parallaxes using the red giant branch relation defined by the Sgr core 
(for which we adopt $[m-M]=16.90$).  
The resulting three dimensional distribution
reveals a flattened, planar alignment with a $<2$ kpc RMS spread, and clear trailing
(predominantly south of the Galactic Plane) and leading (predominantly north of the
Galactic Plane) tidal arms.  
Even without benefit of modeling, the debris arms show clearly that Sgr is orbiting
within a non-precessing plane in an elliptical (approximately
3:1 or 4:1 apo:peri-Galacticon) orbit
of $\sim$14 kpc peri-Galacticon.
Together the tidal arms wrap at least 360$^{\circ}$ around the sky, and overlap at 
even greater length.

To constrain the dynamics of the Sgr-Milky Way interaction further, 
we (e.g., Majewski et al. 2003b, Paper II hereafter) have been collecting radial velocities
of Sgr M giants with the Swope 1-m, KPNO 2.1-m, Yalo 1.5-m and Bok 2.3-m telescopes.
As of July 2003, our sample of M giant velocities, distributed around the Sgr plane, is approaching 
900 M giants.  The $\sim 5$ km s$^{-1}$ precision velocities provide critical input to 
Sgr models, and help compensate for remaining vagaries in the M giant spatial distribution
due to contamination by disk/bulge M giants and random (and perhaps systematic) errors in the
distance scale.

\section{N-body Simulations}

To understand in detail the Sgr interaction with the Milky Way (MW), we (Law et al. 2003b, Paper III)
have been using N-body simulations based on those developed by Johnston et al. (1996, 1999).
Early results of this work are described in Law et al. (2003a).  
For now these models are constructed with a focus on matching the 
structure and kinematics of the Sgr tails; future efforts will aim to understand
the detailed nature of the Sgr core.  Our immediate goal is to use the extensive tidal tail system
revealed by 2MASS M giants as dynamical probes of the MW potential --- its 
size, overall flattening, and lumpiness --- and the global character of the Sgr 
dwarf --- its orbit, mass, and dark matter content.

Sgr is represented by 10$^5$ self-gravitating particles (for
both light and dark matter) distributed according to a Plummer 
model.  
This satellite orbits in
a rigid MW potential represented by a Miyamoto-Nagai (1975) disk, Hernquist (1990)
spheroid, and logarithmic halo --- $\Phi_{halo} = v_{halo}^2 \ln([R^2 + (z/q)^2 + d^2])$ --- 
constrained to fit the established Galactic HI/CO rotation curve
interior to the Solar Circle.
The solar distance to the Galactic Center, $R_{\sun}$, the distance of Sgr, $d_{Sgr}$, 
and the circular speed of the Galaxy, $v_{halo}$, 
are varied
but Sgr's 
radial velocity is fixed at 171 km s$^{-1}$ (Ibata et al. 1997, 
``I97").  The velocity vector of Sgr is constrained to lie
within the orbital plane established by the M giants, and the satellite is evolved through the 
simulated Galactic potential for five orbits using the self-consistent field code of Hernquist
\& Ostriker (1992).

Both full N-body simulations as well as less CPU-intensive, test particle models\footnote{A 
single test particle orbiting in the Galactic potential with the same dynamical 
constraints.} (when appropriate --- e.g., to explore gross orbital properties) 
are used to explore the parameter space of Galactic potential
strength and shape, Sgr orbit,
and Sgr mass; the ranges of some parameters explored are given in Table 1.  Models
are evaluated by their ability to reproduce a set of observed properties of the M giant
data, among the most important and discriminatory including the apo-Galacticon of the  
leading arm, the mean positional, velocity and density trends of the trailing arm, velocity and
positional {\it spreads} of the debris, as well as the amount of precession in the arms (see \S3).
Some degeneracy of parameter combinations yielding reasonable fits to the data is found, 
but fixing $q=1$ and $R_{\sun}=8.5$ kpc yields a ``best-fitting" solution given by the 
``adopted model" in Table 2. 
The latter model (subject to change with further experimentation) 
is characterized by a 0.75 Gyr radial period Sgr orbit with a 14 kpc peri- and 
52 kpc apo-Galacticon, and present space velocity of 326 km s$^{-1}$.  In this model
the observed M giant tidal arms correspond to debris lost on at least 2.5 orbits over
the last $\sim$1.8 Gyr or more.  
For any $R_{\sun}$$\ge$8.5, the MW mass within 50 kpc is restricted to 
3.7-5.1$\times 10^{11}$ M$_{\sun}$, which is slightly smaller than
the recent determination of $5.5 \times 10^{11}$ M$_{\sun}$ 
by Sakamoto, Chiba \& Beers (2003) from an analysis
of the velocities of hundreds of random halo objects.  The model timescales and masses
scale with the potential, while the apo- and peri-Galactica scale with the
M giant distance scale, which was based on the assumed $d_{Sgr}=24$ kpc.

\begin{table}
\caption{Some Parameters in the N-body Models}

\begin{center}
\begin{tabular}{lcc}
\hline
Parameter        &  Range Explored   &  ``Adopted" Model \\
\hline
Solar Galactocentric distance ($R_{\sun}$) & 7.5-9.5 kpc    & 8.5 kpc \\
MW halo circular velocity ($v_{halo}$) & 200-220 km s$^{-1}$ & 210 km s$^{-1}$\\
MW halo flattening ($q$)              &  0.8-1.0                         &  1.0 \\
MW halo softening ($d$)              &  0-50 kpc                       & 9 kpc\\
Sgr distance ($d_{Sgr}$)             & 22-26 kpc                       & 24 kpc \\
Sgr angular momentum &  4309-5427 kpc km s$^{-1}$       & 4788 kpc km s$^{-1}$\\
present Sgr mass &  $4\times10^7-1\times10^9$M$_{\sun}$    &  $3\times10^8$M$_{\sun}$\\
\hline
\end{tabular}
\end{center}
\end{table}

\vskip -0.4in

\section{Constraints on the Halo Flattening}

Because the Sgr orbital plane is {\it not} strictly polar, it should precess in a flattened potential. 
Yet the observed debris plane is remarkably well collimated, suggesting a nearly spherical MW
halo at the distance of Sgr (Ibata et al. 2001, MSWO).  
To quantify this, separate planes are fit to $\sim120^{\circ}$ sections of the leading and
trailing arm corresponding to debris $\sim180^{\circ}$ out of orbital phase
(but up to $300^{\circ}$ separated), and the angle between the planes measured.
Errors reflect the quadrature sum of the errors in fitting the two planes.
Table 2 gives the results of this comparison for the observed M giant debris 
and for closely matching simulations varying only by the degree of 
flattening in the halo potential.

\begin{table}
\caption{Amount of Precession for Different Halo Flattenings}

\begin{center}
\begin{tabular}{lcc}
\hline
Data Set               &   $q$   &  Precession (degrees) \\
\hline
2MASS M giants &  ...        & $1.7\pm2.4$ \\
simulation            & 1.00    & $2.2\pm1.6$ \\
simulation            & 0.95    & $3.5\pm1.7$ \\
simulation            & 0.90    & $5.6\pm1.4$ \\
simulation            & 0.85    & $10.7\pm1.0$ \\
\hline
\end{tabular}
\end{center}
\end{table}

As may be seen, the distribution of observed M giants is fully consistent with a $q=1$, spherical 
halo potential,
although a 1$\sigma$ error permits a slight flattening of the halo ($q=0.95$).  The constraint
on halo flattening offered by the Sgr tidal arms will strengthen as the length over which 
the arms are traced increases, both by M giants (possible when we verify the radial velocity membership
of potential M giants at even greater separations from the core --- an observing program in progress)
and by use of {\it older} stellar tracers that can track the arms beyond the length 
that can be traced with the $\sim$ 2-3 Gyr old Sgr M giants.  Based on these results, we have
adopted a $q=1$ halo potential in our simulations.

\section{Constraints on the Bound Sgr Mass and $M/L$}

The $M/L_V$ of Sgr has previously been found to be very large --- of order
50 (I97) to 100 (Ibata \& Lewis 1998).  However, MSWO showed that the
M giants of the Sgr core can be fit by a King model with much larger 
core and limiting radii than previously adopted. 
Inserting these radii and the 11.4 km s$^{-1}$ central velocity dispersion (I97) 
into the standard King (1966) $M/L$ methodology that assumes virial equilibrium
yields a $4.9\times10^8$M$_{\sun}$ bound Sgr mass 
that drops $(M/L_V)_{tot}$ to 25.  However, as MSWO 
point out, there is little reason to believe even this $M/L$ because it is very unlikely that 
the Sgr tidal radius corresponds to the measured 12.6 kpc 
semi-major limiting radius.  For example, adopting this as a tidal 
radius yields an absurdly discrepant $1.6\times10^{11}$ M$_{\sun}$ 
Sgr mass from the Roche tidal limit,
$m_{Sgr} = [2M_{MW}(R_{GC})] [r/R_{GC}]^3$. Even adopting the 4.4 kpc {\it minor} 
axis limiting radius still yields $m_{Sgr} = 6.9\times10^{9}$ M$_{\sun}$ and $M/L_v \sim 343$.
The tidal radius must be much smaller, especially if we are
to explain how 2-3 Gyr old M giants, presumably formed in the central few kpc of Sgr, could so
quickly have escaped across the tidal boundary into tidal arms of similar
dynamical age.  With no clear physical markers of a tidal radius in the spatial distribution 
of Sgr core stars, we must resort to alternative means to estimate its mass.

Although we do not attempt to model the Sgr core in detail, we can use the 
coherence of the Sgr stream to estimate the mass of the disintegrating parent core, 
since larger mass bodies produce commensurately ``hotter", wider debris trails.
The simulated dwarf that best fits the spatial and velocity 
width of the streams has a $3\times10^8$M$_{\sun}$
mass within a semi-major tidal radius of about $r_{\rm tide} = 3.5$ kpc, 
which gives $(M/L_V)_{tot} =21$ (adopting the Sgr $L_V$ within this radius).
This may be regarded as an {\it upper limit} to the Sgr core mass (see \S5).

When the Sgr arms can be mapped accurately over even greater lengths,
it may be possible to map the degree to which {\it dynamical friction} has acted on the Sgr core, 
and thereby derive yet another, independent estimate of the Sgr mass.
We have explored dynamical friction models as a means to explain one apparent discrepancy
between the model and M giant velocities in the nearest sections of the Sgr leading arm, however
no satisfactory results have been obtained that do not imply a huge recent mass loss and
extreme former Sgr core mass.

\section{Constraints on the Lumpiness of the Milky Way Halo}

Current CDM models for the formation of structure in the universe
predict that MW-like galaxies should contain substantial halo substructure 
at current epochs as a result of the accretion of thousands of subhalos over a Hubble time
(e.g., Navarro et al.\ 1997).
Because the MW currently has only eleven known {\it luminous} satellite galaxies, 
it is commonly
held that the bulk of the subhalos must be made up of pure dark matter (see Klypin et al. 1999, 
Moore et al. 1999).  If so, these dark matter lumps should make their presence known
through the heating of dynamically cold, {\it luminous} stellar systems, like
tidal tails 
(Moore et al. 1999, Font et al.\ 2001, Johnston, Spergel \& Haydn 2002, Ibata et al. 2001, 2002).

The present velocity and spatial dispersion of the Sgr tails reflect both 
the initial dispersion of tidally released debris
as well as subsequent heating of that debris imparted by encounters with large halo masses.
Attributing all of the dispersion to one or the other of these phenomena provides upper
limits to the effects of each.  We (Paper II) have measured a $10.4\pm1.5$ km s$^{-1}$ velocity dispersion
over a $>$100$^{\circ}$ expanse of 
the trailing Sgr arm, a dispersion nearly equivalent that of the Sgr core.
If we attribute all of this dispersion to the Sgr central mass, then we derive an upper limit
to that mass of 3$ \times 10^8$ M$_{\sun}$ and $M/L_v < 21$ (as described above).
However, if any of the velocity dispersion in the tidal arms
is attributable to heating by subhalos, the implied bound mass of Sgr decreases.
While the velocity dispersion of the trailing Sgr arm actually is observed to increase slightly
with distance from the Sgr core (Paper II), it is not yet clear 
whether this is the signature of subhalo heating or simply the fact that
the bound Sgr mass was larger in the past.

Johnston et al. (2002, ``JSH") give a prescription for a tidal debris ``scattering index", $B$,
that measures position and velocity perturbations  of tidal arm stars
induced by lumps in the halo under the assumption of an initial zero velocity dispersion
tidal debris population (thus the index provides an
{\it upper limit} to the scattering for real debris).  
When applied to trailing arm M giants with 25-90$^{\circ}$ separation from the Sgr core
we obtain $B=0.031$, a ``colder" result than the 0.037 value JSH obtained for
presumed Sgr carbon stars.  The new, smaller $B$ is consistent with JSH simulations
of heating in a smooth halo containing just one LMC-like (mass and orbit) lump;
however, some realizations of lumpier halos in the JSH analysis are
not inconsistent with the degree of scattering observed here.  Unfortunately,
this reflects a vagary of this type of halo probe: Dynamical heating in CDM halos
tends to be dominated by the most massive lumps.  

Nevertheless, still tighter constraints --- in the direction of making
lumpy halos even less likely --- may derive
from future observations and modeling of tidal streams.  
For example, accurately determining the zero-age dispersion of Sgr debris will
make it possible to remove this contribution from any dispersion by heating.
The study of initially colder streams, e.g., from globular clusters, 
would place even stricter constraints on the lumpiness of the halo.

\medskip
We acknowledge funding from NASA, the National Science Foundation, 
and the David and Lucile Packard Foundation.


\begin{references}

\reference Bessell, M.~S.~\& Brett, J.~M.\ 1988, \pasp, 100, 1134

\reference Font, A.~S., Navarro, J.~F., Stadel, J. \& Quinn, T. 2001, \apjl, 563, L1

\reference Hernquist, L. 1990, ApJ, 356, 359

\reference Hernquist, L. \& Ostriker, J. 1992, ApJ, 386, 375

\reference Ibata, R.~A, Gilmore, G. \& Irwin, M.~J. 1994, Nature, 370, 194

\reference Ibata, R.~A., Lewis, G.~F., Irwin, M.~J. \& Quinn, T. 2002, \mnras, 332, 915

\reference Ibata, R., Lewis, G.~F., Irwin, M., Totten, E. \& Quinn, T.\ 2001, \apj, 551, 294

\reference Ibata, R.~A., Wyse, R.~F.~G., GIlmore, G., Irwin, M.~J. \& Suntzeff, N. B. 1997, AJ, 113, 634
(I97)

\reference Johnston, K.~V., Hernquist, L. \& Bolte, M. 1996, ApJ, 465, 278

\reference Johnston, K.~V., Majewski, S.~R., Siegel, M.~H., Reid, I.~N. \& Kunkel, W.~E. 1999, AJ, 118, 1719

\reference Johnston, K.~V., Spergel, D.~N. \& Haydn, C. 2002, \apj, 570, 656 (JSH)

\reference King, I.~R.\ 1966, \aj, 71, 64

\reference Klypin, A., Kravtsov, A.~V., Valenzuela, O. \& Prada, F.\ 1999, \apj, 522, 82

\reference Law, D.~R., Johnston, K.~V. \& Majewski, S.~R. 2003b, in prep. (Paper III)

\reference Law, D.~R., Majewski, S.~R., Skrutskie, M.~F. \& Johnston, K.~V. 2003a, in
ASP Conf. Ser. Vol., Tidal Tails and Galactic Satellites, eds. D. Martinez-Delgado \&
F. Prada, in press (astro-ph/0309567)

\reference Majewski, S.~R., Skrutskie, M.~F., Weinberg, M.~D. \& Ostheimer, J.~D. 2003a, ApJ, in press
(astro-ph/0304198; MSWO)

\reference Majewski, S.~R., et al. 2003b, AJ, submitted (Paper II)


\reference Miyamoto, M. \& Nagai, R. 1975, PASJ, 27, 533

\reference Moore, B., Ghigna, S., Governato, F., Lake, G., Quinn, T., Stadel, J. \& 
Tozzi, P.\ 1999, \apjl, 524, L19

\reference Navarro, J.~F., Frenk, C.~S. \& White S.~D.~M. 1997, \apj, 490, 493


\reference Sakamoto, T., Chiba, M. \& Beers, T.~C.\ 2003, \aap, 397, 899

\reference Sparke, L.~S. 2002, in The Shapes of Galaxies and Their Dark Matter Halos, 
ed. P. Natarajan, (Singapore: World Scientific), p. 178

\end{references}
\end{document}